\documentclass[12pt]{amsart}
\usepackage{amsmath,amsthm}

\def\pmatrix{\left(\begin{matrix}}
\def\endpmatrix{\end{matrix}\right)}

\def\Sp{\operatorname{Sp}}

\def\Z{{\mathbb Z}}
\def\F{{\mathbb F}_2}
\def\C{{\mathbb C}}

\def\de{\delta}

\def\t{\theta}

\def\e{\varepsilon}

\def\M{{\mathcal M}}

\def\H{{\mathcal H}}

\def\tch#1#2{{\left[\begin{matrix}#1\\ #2\end{matrix}\right]}}
\def\tt#1#2{{\t\tch{#1}{#2}}}

\theoremstyle{plain}
\newtheorem{thm}{Theorem}
\newtheorem{lm}[thm]{Lemma}
\newtheorem{prop}[thm]{Proposition}
\newtheorem{cor}[thm]{Corollary}

\theoremstyle{definition}

\newtheorem{rem}[thm]{Remark}
\begin{document}
\title{ Remarks on Superstring  amplitudes in higher genus}
\author{Riccardo Salvati Manni}
\address{Dipartimento di Matematica, Universit\`a "Sapienza'',
Piazzale A. Moro 2, Roma, I 00185, Italy}
\email{salvati@mat.uniroma1.it}
\date{\today}
\begin{abstract}
Very recently, Grushevsky  continued D'Hoker and Phong's   program of finding the chiral superstring measure from first principles by constructing modular forms satisfying certain factorization constraints. He  has proposed an ansatz in genus 4 and  conjectured a possible formula for the superstring measure in any genus, subject to the condition that certain modular forms admit holomorphic roots. In this note we want to give some evidence that Grushevsky 's approach  seems to be  very fruitful.  
\end{abstract}
 \maketitle
\section{Introduction}
D'Hoker and Phong  managed, in a sequence of papers,  to compute the genus 2 superstring measure from first principles in terms of modular forms   cf. \cite{DHP1}, \cite{DHPa}, \cite{DHPb},\cite{DHPc}  and verified some corresponding results such as the vanishing  cosmological  constant, 2-3- point scattering amplitudes et cetera.   
Since a derivation in higher genus from first principles
appears prohibitively difficult at the present time, as an
alternative approach, they suggested looking for ansatze
satisfying factorization constraints consistent with the
genus 2 formula, and in particular a certain set of conditions
for an eventual genus 3 ansatz cf .\cite{159} and \cite{ 182}.\smallskip

Cacciatori, Dalla Piazza, and van Geemen, cf. \cite{CDPvG}  proposed an ansatz for the chiral superstring measure in genus 3 in which they relaxed one of the  conditions  imposed by D'Hoker and Phong.
This relaxing appears necessary otherwise the solution to the constraints of D'Hoker and Phong does not exist. Very recently Grushevsky rewrote the ansatz of D'Hoker and Phong in genus 2, and of Cacciatori, Dalla Piazza, and van Geemen in genus 3 in terms of modular forms associated to isotropic spaces of theta characteristics. This allows  him  to give  a straightforward generalization of the chiral superstring measure to higher genera, which for genus 4 is an appropriate holomorphic modular form . For higher genera he conjectures a possible ansatz, satisfying the factorization constraints,but involving  holomorphic roots. of suitable monomials in the theta constants.
In this note we want to give some evidence that Grushevsky's approach  seems to be the more fruitful.  
 In fact we shall consider   properties of products of theta constants with characteristics forming   even cosets    of   isotropic subspaces. We prove  that in genus 5, Grushevsky's  ansatz, that involves square roots of theta constants is holomorphic on $\M_5 (1,2)$.
 Another interesting   results that we will get is to
 verify, without any tedious computation,  that  Grushevsky's proposed ansatz satisfies further physical constraints, for example that it is easy to prove the vanishing  of  cosmological constant when $g\leq 4$.

 \section{Notations, definitions and basic facts}
 
We denote by $\H_g$ the Siegel
upper half-space of symmetric complex matrices with positive-definite imaginary part, called period matrices. The  action of the symplectic group $\Sp(g,\Z)$ on $\H_g$ is
given by
$$
  \pmatrix A&B\\  C&D\endpmatrix\circ\tau:= (C\tau+D)^{-1}(A\tau+B)
$$
where we think of elements of $\Sp(g,\Z)$ as of consisting of four
$g\times g$ blocks, and they preserve the symplectic form given in
the block form as $\pmatrix 0& 1\\ -1& 0\endpmatrix$. 

For a period matrix $\tau\in\H_g$,  $z\in \C^g$  and $\e,\de\in \F^g$ (where $\F$ denotes the abelian group $\Z/2\Z=\lbrace 0,1\rbrace$) 
the associated theta function with characteristic $m=[\e, \de]$ is
$$\t_m(\tau, z)=
  \tt\e\de(\tau,z)=\sum\limits_{n\in\Z^g}\exp(\pi i (\tau[
  (n+\e/2)]+ 2t(n+\e/2)'( z+\de/2)).
$$
Here  we denote by $X'$  taking the transpose of $X$ and   $A[X]:=X'AX$
As a function of $z$, $\t_m(\tau, z)$ is odd or even depending on whether
the scalar product $\e\cdot\de\in\F$ is equal to 1 or 0,
respectively. \smallskip

Theta constants are restrictions of theta
functions to $z=0$. All odd theta constants vanish identically in
$\tau$.  
 We define the level subgroups of $\Sp(g,\Z)$ as follows:
$$
  \Gamma_g(n):=\left\lbrace \gamma=\pmatrix A&B\\ C&D\endpmatrix
  \in\Gamma_g\, |\, M\equiv\pmatrix 1&0\\
  0&1\endpmatrix\ {\rm mod}\ n\right\rbrace
$$
$$
  \Gamma_g(n,2n):=\left\lbrace \gamma\in\Gamma_g(n)\, |\, {\rm
  diag}(AB')\equiv{\rm diag} (CD')\equiv0\ {\rm mod}\
  2n\right\rbrace.
$$
Here  $diag (X)$ means the vector consisting of the diagonal 
entries of the square matrix $X$. 
These are normal subgroups of $\Sp(g,\Z)$  for $n$ even. 

We shall write $\Gamma_g$ for 
$$\Gamma_g(1)=\Sp(g,\Z)$$

Let $k$ be a positive integer and $\Gamma$ be a  subgroup of finite index in $\Gamma_g$ 
A multiplier system of weight $k/2$ for $\Gamma$ is 
a map $v:\Gamma\to \C^*$, such that the
map
$$
  \gamma\mapsto v(\gamma)\det(C\tau+D)^{k/2}
$$
satisfies the cocycle condition for every $\gamma\in\Gamma$ and
$\tau\in\H_g$.

With these notations,  we say that
a holomorphic function $f$ defined on $\H_g$ is  a modular form 
of weight $k/2$ with respect to $\Gamma$ and $v$ if

$$
  f(\gamma\circ\tau)=v(\gamma)\det(C\tau+D)^{k/2}f(\tau)\quad
  \forall\gamma\in\Gamma,\forall\tau\in\H_g.
$$
and if additionally $f$ is holomorphic at all cusps when $g=1$\smallskip
 We denote by $[\Gamma, k/2, v]$ the  vector space of such functions.
We omit the multiplier if it is trivial.
\smallskip

The full symplectic
group acts on theta constants with characteristics as follows:
$$
  \theta_{\gamma\cdot m} (\gamma\cdot\tau)=$$
  $$\kappa(\gamma) \exp((1/4)\pi i (B'D[\e]-2\e'B'C\de+A'C[\de] -2diag(AB')(D\e-C\de))    )$$
  $$det(C\tau+D)^{\frac{1}{2}}\theta_m
  (\tau),
$$

\indent 
where $\kappa(\gamma)$ is an eighth root of unity, and the action
on the characteristic is
$$
\gamma\cdot m=  \gamma\pmatrix \e\cr \de\endpmatrix :=\pmatrix D&-C\cr
  -B&A\endpmatrix \pmatrix \e\cr \de\endpmatrix+ \pmatrix {\rm diag}(CD')\cr {\rm diag}(AB')\endpmatrix
$$
 
where the addition in the right-hand-side is taken in $\F^{2g}$.

We observe at  a certain point we shall consider the addition in the right-hand-side is taken in $\Z^{2g}$

In \cite{I}  it is proved that theta constants are modular forms of weight $1/2$ with respect to $\Gamma_g( 4,8)$ . 

The action of $\Gamma_g(2)$ on
the set of theta constants produces a multiplier.
Let $M=(m_1, m_2,\dots, m_k)$ be a sequence of characteristics,
we set
$$P(M)=\prod_{i=1}^k\t_{m_i}$$
For  $k=2r$   even,  in  the previous discussion we get that that $P(M)$ is a modular m with respect to $\Gamma_g (2)$ and  a suitable  character. From \cite{tc} , we know that the character is trivial if the matrix
 $M$ satisfies the
 congruences
 $$M'M\equiv r\pmatrix 0&1_g\\ 1_g&0\endpmatrix\quad{\rm mod}\, 2$$
 $$ diag(M'M) \equiv 0 \quad{\rm mod}\, 4.$$
For all characteristic $m=[\e, \de]$ and $n=(\e_1, \de_1)$, we set
$$e(m,n)= (-1)^{\e\cdot\de_1 +\e_1\cdot\de}.$$ 

We say that a  subspace  $V\subset\F^{2g}$ is isotropic
if $e(m,n)=1$ for all $m,n$ in $V$. A maximal isotropic space has  dimension $g$. Each isotropic subspace has even cosets , i.e cosets all of which entries are even characteristics. For details we refer to
\cite{SM}. Let $V+m$ be an even coset of  an isotropic  $j$ dimensional space. Let us assume that
$2^j s=2^k\geq 8$, thus the form
 $P(V+m)^s$ belongs to $[\Gamma_g(2), 2^{k-1}].$
 for details we refer to\cite{Ch} and \cite{SM}.

One can thus study the orbits of characteristics or sets of
characteristics under the symplectic group action. This was done in \cite{Ja}
\cite{TN}, where, in particular,   it can be found that $\Sp(g,\F)$ acts transitively on the
even cosets of    isotropic  $j$ dimensional spaces.

Moreover    $\Gamma_g(1,2)/\Gamma_g(2)$ acts transitively on  even isotropic  $j$ dimensional spaces. As consequence of these facts we get the following

\begin{lm} Let  us assume that  $8$ divides $2^i s$ and $g\geq i$, thus the form
$$S_{i,s}^g=\sum_{V+m} \e_{V+m} P(V+m)^s(\tau)$$
belongs to $[\Gamma_g,  2^{k-1}]$
providing that we are summing over  all even cosets of all $i$ dimensional even isotropic spaces and $\e_{V+m}$ are suitable signs .
\end{lm}

{\it Proof}   Since $\Sp(g,\F)$ acts transitively on the  set of even cosets of  isotropic  spaces of the same dimension, we  will sum over the  associated  monomials in the theta constants using the transformation formula. Thus we can  start from  a fixed sequence $M$, corresponding to the vectors of a even totally isotropic space $V$.  let $ G_V$ be the stabilizer of $V$ in  $\Sp(g,\F)$, thus from the transformation formula,\cite{TN} we get that
$$\frac{1}{|G_V|}\sum_{\gamma\in G} det(C\tau+D)^{-2^{k-1}}P(M)^s(\gamma\cdot\tau)=\sum_{V+m} \e_{W+m} P(W+m)^s(\tau)$$
 with$$\e_{W+m}=\exp((1/4)\pi i (tr(  AB'[ M_1]+ CD' [M_2])$$
 $$\exp((1/2)\pi i (tr(CB'M_1M_2'+(\gamma^{-1}M)_1' (N-\gamma^{-1}M)_2)$$
 
 Here we denoted with $W+m$ the even coset obtained by $V$ acting with $\gamma^{-1}$.
 This means that if
 $M=\pmatrix M_1\\ M_2\endpmatrix $  is the sequence of vectors in $V$, $\gamma^{-1}M$ is the sequence  of vectors in  $W+m$ with coefficients in $\F$.  Here $N$ is a sequence of vectors in  $\Z^{2g}$ induced by the action of $\Gamma_g $ on $M$  with  the addition  in $\Z$.
 Obviously we have
 $N\equiv \gamma^{-1}M \quad {\rm mod}\, 2.$
 
By construction this form belongs  necessarily to $[\Gamma_g,\,   2^{i-1}s]$.
A priori it could be identically zero.
The congruences satisfied by the matrix $sM$, imply that $\e_V=\pm 1$.\smallskip

 We can refine the previous result  proving an interesting result and a consequent corollary that result to be fundamental in \cite{GR}

\begin{prop} Let  us assume that  $16$ divides $2^i s$  and $g\geq i$,
$$S_{i,s}^g=\sum_{V+m}   P(V+m)^s(\tau)$$
belongs to $[\Gamma_g, \,  s2^{j-1}]$
provided that we are summing over all even cosets of $i$ dimensional  isotropic spaces 

\end{prop}
{\it Proof}\,  We have to prove that in these cases $\e_V=1$.
In fact the matrix $sM$ satisfies the
 congruences
 $$^tMM\equiv \pmatrix 0&0\\ 0&0\endpmatrix\quad{\rm mod}\, 4$$
 $$ diag(^tMM) \equiv 0 \quad{\rm mod}\, 8.$$
 An immediate computation shows that  from the previous formula  we have   possibly  non trivial term only when $s=1$ ( hence $dim V\geq 4$). This is
 
$$\exp((1/2)\pi i (tr( (\gamma^{-1}M)_1' (N-\gamma^{-1}M)_2)$$
Now to compute this term we can use a set of generators of the group $\Sp(g,\F)$ or just observe that in  \cite{Ch} the case of i=4 for the action of $\Gamma_g$  was considered.
In this case appear all even coset of four dimensional totally isotropic spaces and  the coefficients are all $1$.  Since an even coset  of a   isotropic subspace  $V$of dimension $j+1$  is  union of  two even cosets of  a suitable  isotropic subspace $W$ of dimension $j$, by induction we see that the 
above term is always 1. In fact  if $$V+m= ( W+m)\cup(W+n)$$ we  have that 
$$\e_{V+m}=\e_{W+m}\e_{W+n}=1.$$

As consequence of the above proposition we get the following corollary stated  also in\cite{GR}

\begin{cor} Let  us assume that  $16$ divides $2^i s$,  
$$P_{i, s}^g (\tau)= \sum_V   P(V)^s(\tau)$$
belongs to $[\Gamma_g(1,2),\, 2^{i-1}s]$
provided that we are  summing over $j$ dimensional even isotropic spaces 

\end{cor}

\begin{rem} A similar statement holds  even if $8$ divides $2^i s$, but we will get some signs.
To avoid confusion, from now on we 
assume that 16 divides $2^is$.
\end{rem}

We observe that $P_{i, s}^g (\tau)$ can be considered  also as the partial sum of  $S_{i, s}^g (\tau)$
where we restrict  our computation to all even cosets of $j$ dimensional even isotropic spaces 
containing the $0$ characteristic.
 Thus similarly to $P_{i, s}^g (\tau)$, for every even characteristic $m$, we can define
 $P_{i, s}^g[m] (\tau)$ where we take the sum over even cosets   cosets containing the characteristic $m$. Obviously
$P_{i, s}^g[0] (\tau)=P_{i, s}^g (\tau)$
and
$$2^iS_{i, s}^g (\tau) =\sum_{m\, even} P_{i, s}^g[m] (\tau).$$

We observe that the coefficient $2^i$ of   $S_{i, s}^g$ is due to the fact that in the summation in the right  hand side   an even coset of  a $i$ dimensional subspace appear with  multeplicity $2^i$ , i.e. the  number of vectors in the coset.  \bigskip

Obviously $P_{i, s}^g[m] (\tau)$ are modular form with respect to a subgroup of $\Gamma_g$ conjugated to  $\Gamma_g (1,2)$. We recall that  $\Gamma_g (1,2)$ is not normal in  $\Gamma_g$.

\section{ Grushevsky's results}
In this short  section we recall the main results obtained in \cite{GR}. Obviously we can assume  that  $16$ divides $2^i s$. All details can  be found in the above cited paper.\smallskip

The modular forms $P_{i, s}^g (\tau)$ has the following remarkable property

\begin{prop} The modular forms  $P_{i, s}^g$ restrict to the locus of  block diagonal period matrices $\H_k\times\H_{g-k}$ as follows
$$P_{i, s}^g = \sum_{0\leq n,m\leq i\leq n+m}N_{n,m;i} P_{n, 2^{i-n}s}^k  P_{m, 2^{i-m}s}^{g-k}$$
with
$$N_{n,m;i}=\prod_{j=0}^{n+m-i-1}\frac{(2^n -2^j)(2^m-2^j)}{2^{n+m-i }-2^j}$$
\end{prop}

These forms are the basic bricks for an easy writing of the  low genus superstring scattering amplitudes proposed  by D'Hoker and Phong \cite{DHP1} for genus 2 and by Cacciatori, Dalla Piazza, and van Geemen \cite{CDPvG} for genus 3. In fact in \cite{GR}  has been proved
\begin{thm}
For $g \leq 4$  the  function
$$
 \Xi^{(g)}[0]:=\frac{1}{2^g}\sum\limits_{i=0}^g (-1)^i2^{\frac{i(i-1)}{2}}P_{i, 2^{4-i}}^{g}
$$
is a modular form of weight 8 with respect to $\Gamma_g(1,2)$, such that its restriction to $\H_k\times\H_{g-k}$ is equal to $\Xi^{(k)}[0]\cdot\Xi^{(g-k)}[0]$.
\end{thm}
Thus these are   natural candidates for the  superstring measures.

Moreover in  \cite{GR}, it is shown that the above statement  holds for g>4 as well, up to a possible inconsistency in the choice of the roots.

\section{Evidences for the ansatz}

In this section we shall give 
evidence for these  natural candidates verifying that the proposed ansatz satisfies further physical constraints.  \smallskip
 
Similarly to the $P's$ case  we define
$$\Xi^{(g)}[m]=\frac{1}{2^g}\sum\limits_{i=0}^g (-1)^i 2^{\frac{i(i-1)}{2}}P_{i, 2^{4-i}}^{g}[m]$$

Summing up we get
\begin{lm} 
$$ \Xi^{(g)}=\sum_m \Xi^{(g)}[m] =\frac{1}{2^g}\sum\limits_{i=0}^g (-1)^i 2^{\frac{i(i+1)}{2}}S_{i, 2^{4-i}}^g$$

\end{lm}

Showing that the cosmological constant vanishes is equivalent to show
 that  $\Xi^{(g)}$ is identically zero on  the moduli space of curves $\M_g$ This has been verified for genus 2 in \cite{DHP1} and  for genus 3 in \cite{CDPvG}. We shall prove it in genus 4. \smallskip
 
 This is a consequence of remarkable formulas proved  around thirty years ago \cite{Ch}. In fact in the above cited paper Igusa introduced modular forms of weight 8 related to $\Gamma_g$ and relations among them.

 \begin{lm}We have
$$(2^{2g}-1)S_{0,16}= 6S_{1,8}+24S_{2,4}$$
$$(2^{2g-2}-1)S_{1,8}= 18S_{2,4}+168S_{3,2}$$
$$(2^{2g-4}-1)S_{2,4}= 42S_{3,2}+840 S_{4,1}$$
for $g\geq 2,3,4$ respectively

\end{lm}

As  immediate consequence of the lemma we get the following

\begin{thm}
The modular form 
$\Xi^{(g)}$ is identically zero on $\M_g$ 
 for $g\leq 4$
\end{thm}

{\it Proof} We shall prove only the case $g=4$. The other are similar and easier
According to our previous discussions we have
$$2^4\Xi^{(4)}=S_{0,16}-2S_{1,8}+8S_{2,4}-64S_{3,2}+1024S_{4,1}$$
From previous lemma   we get
$$\Xi^{(4)}= 2^2\cdot3\cdot  7^{-1}(15S_{0,16}-2S_{1,8})$$
According to 
theorem 1 in \cite{Ch}, we get that

$$\Xi^{(4)}=-2^8 \cdot 3^3\cdot 5  J(\tau)$$

Here $J(\tau)$ is  Schottky's polynomial that vanishes along $\M_4$.\bigskip

 We end our discussion, considering  the proposed   ansatz for the superstring measure in any genus,
in the genus $5$ case .
 In this case the theorem involves  square root  of modular forms.

 $$
 \Xi^{(5)}[0]:=\frac{1}{2^5}\sum\limits_{i=0}^5 (-1)^i 2^{\frac{i(i-1)}{2}}P_{i. 2^{4-i}}^{5}
$$
 involves the term $P_{5, 1/2}^{5}$.
 Now the problem of the definition of  square roots  of  theta constants on suitable covering of $\M_g$
 has been settled in \cite{Ts}.  First of all let us define  $\M_g (n, 2n)$ as the inverse image of  $\M_g$ in $\H_g/\Gamma(n,2n)$ under the standard projection map.    $\sqrt{\t_m}$ can be considered as a modular form on $\M_g(4,8)$,  in fact the  divisor of theta constants are  tangent to $\M_g(4,8)$ and this allows to have square roots.\smallskip
The main problem involves the factor of automorphy of the transformation formula, in fact for any subgroup of finite index in $\Gamma_g$  the  $\sqrt{\t_m}$ have diffent multipliers.
 These obstruction can be eliminated in some special case , in fact  as consequence of relation of Schottky type in \cite{Ts} it  has been proved 
 
\begin{lm} Let $M$ be a   isotropic   3 dimensional space, then for all ( even) cosets $M+m$
$$\sqrt {P(M+m)}$$
belong to the  same automorphy  factor in  $\M_g(4,8)$
\end{lm}
Now  let us assume $g=5$.  Even cosets of maximal isotropic spaces are only isotropic spaces. Let $V$  be a maximal 5  dimensional even isotropic space,
 Since 
 $V= M\cup M+n_1\cup M+n_2\cup M+ n_3$
 for  a suitable  3 dimensional,   isotropic space $M$.
 Hence the factor of automorphy of  the term $\sqrt {P(V)}$ appearing in $P_{5, 1/2}^{5}$  is the fourth power of $\sqrt {P(W)}$ where $W\subset V$ is a three dimensional isotropic space. First of all we want to prove that  all  $\sqrt {P(V)}$, $V$  even maximal isotropic space have the same factor of automorphy.
 This can be reduced to a combinatorial  
 
 \begin{lm} Let $V, Z$ be even maximal isotropic space in genus 5
 then there exist  a sequence  of   even maximal isotropic space $V_0=V, V_1,\dots, V_n=Z$
 and a sequence of even 3 dimensional  isotropic space $M_1,\dots, M_n$ such that
 $$M_i\subset V_{i-1}\cap V_i $$
 \end{lm}
 {\it Proof}  Let $x_1, \dots x_{10}$ be the coordinates of $\F^{10}$, without any loss of  generality ( all the other cases are easier)  we can assume that 
 $ V=V_0$ is defined by $\{ x_1=x_2=x_3=x_4=x_5=0\}$ and $ Z=V_n$  is defined by $\{x_6=x_7=x_8=x_9=x_{10}=0\}$. Let $M_1$ be the subspace of $V_0$   defined by $\{x_6=x_7=0\}$.\smallskip

 \noindent $M_1$ is obviously contained in the subspace $ V_1$ defined by $\{x_3=x_4=x_5=x_6=x_7=0\}$.  
   Let $M_2$ be the subspace of 
   $V_1$ defined by $\{x_8=x_9=0\}$.  $M_2\subset V_2$
 $V_2$  is defined by $\{ x_5=x_6=x_7=x_8=x_9=0\}$
 Now $M_3\subset V_2$ $\{x_{10}=x_4=0\}$
  Hence  $M_3\subset V_3 =Z$
   As  consequence of the above lemma we have
   
    \begin{thm} $P_{5, 1/2}^{5}$  is  a modular form relative  to $\Gamma_5(1, 2)$, once we restrict to the $ \M_5(1, 2)$, hence  $\Xi^{(5)}[0]$ is well defined  on $ \M_5(1,2)$.
        \end{thm}
    
   {\it Proof}  We restrict to the  inverse image of $\M_g$ in $\H_g$. From the previous lemma, it follows that all terms  appearing in $P_{5, 1/2}^{5}$ have the  same factor of automorphy w.r.t $\Gamma_5(4,8)$.  In  fact the factor of automorphy of  $P(V)$ is the  fourth power of the factor of automorphy  of  $P(M_1)$. But this is the  same factor of  automorphy of  $P(V_1)$.
   Now we  can iterate the process .  The factor of automorphy of  $P(V_1)$ is the  forth power of the factor of automorphy  of  $P(M_2)$. But this is the  same factor of  automorphy of  $P(V_2)$ an so on.
   \smallskip
   
   By transformation formula this implies that the factor of automorphy is always the same also at the $\Gamma_5(2)$ level. The action of the group $\Gamma_5(1,2)/\Gamma_5(2)$ produces the modular form $P_{5, 1/2}^{5}$.
   Hence $P_{5, 1/2}^{5}$ is a modular form wrt $\Gamma_5(1,2)$ with possibly a character induced by the  square root  .\smallskip
   
We observe that the previous lemma holds  even if $V$ and $Z$ are even  4  dimensional maximal isotropic space in genus 5. Now we can repeat the same argument with terms
 $\sqrt {P(W)}$ where $W$ is    even 4  dimensional  isotropic space in genus 5 .
 Hence $\sqrt {P(W)}$ and  $\sqrt {P(W+m)}$ have the same factor of automorphy, here $W+m $ is  an   even coset. 
Since we  have always a decomposition
$V=W\cup W+m$, we have the character appearing in the transformation formula for $\sqrt{ P(V)}$ is the square of the character appearing in the transformation formula for $\sqrt {P(W)}$. From Lemma 1, we  know that $P(W)$ has trivial character, hence $\sqrt {P(V)}$ has trivial character.\medskip

Since similar results hold for $\Xi^{(5)}[m]$, we get 
 
\begin{cor}
 $\Xi^{(5)}$ is well defined  on $ \M_5$
\end{cor}

\section{ Further remarks}
Because of Grushevsky' s result , it seems reasonable to  consider also the modular forms
$S_{i, s}^g (\tau)$ without  any restriction on the weight. In this section we  shall show that they
have   some nice properties as modular forms. For doing this we need to introduce the Siegel $\Phi$-operator.  
For $f:H_g\to \C$ 
holomorphic we set formula
$$\Phi(f)(\tau_1)=\lim_{ \lambda \longrightarrow + \infty}f \pmatrix \tau_1 &0\\ 0&i\lambda\endpmatrix$$
 for all $\tau_1\in\H_{g-1}$. In particular this operator maps
 $[\Gamma_g, k, ] $ to $[\Gamma_{g-1}, k]$ and   $[\Gamma_g (n, 2n) , k ] $ to $[\Gamma_{g-1} (n, 2n), k] $. This operator has a relevant importance in the theory of modular forms, we refer to \cite{I} or \cite{Fr} for details.  In the case of the full modular  group , a cusp form is a modular form that is in the  kernel of the $\Phi$ operator. In the case of  subgroup of the modular group, a  modular form is a cusp form if all its conjugate  with respect to the $\Gamma_g $ action  are  in the kernel of the $\Phi$ operator. For any of such groups , 
we shall denote by $[\Gamma,\, k]_0$ the subspace of cusp forms. \smallskip

 \noindent The image $\Phi(\theta_m)$ can be easily 
computed. I 
Indeed  if $m=[\e,\de]$ be an even characteristic  with $\e=(\e', \e_g)=(\e_1,\e_2,\dots\e_g)$ and similarly  $\de=(\de', \de_g)=(\de_1,\de_2,\dots\de_g) $ , then we have
 
 $$\Phi(\t_m)(\tau_1)=\begin{cases}
  \t_{m'}(\tau_1)\,\,\,  if \,\,\, \e_g=0,\\
    0\,\, if \,\, \e_g=1\
  \end{cases}$$

 Here $m'\in \F^{2g-2}$ is the characteristic obtained by $m$ deleting  the  entries $\e_g $ and $\de_g$. \smallskip 
 
We want to apply the $\Phi$ operator to the modular forms $S_{i, s}^g (\tau)$.  
  As consequence of simple computations or even as consequence of Grushevsky's result, we get
   \begin{lm} Let us   assume  that $0\leq i\leq g$ and  $16$ divides $2^i s$, then
   
   $$\Phi (S_{i, s}^g (\tau))= 2^{i+1} S_{i, s}^{g-1}+ S_{i-1, 2s}^{g-1}$$
    when $0<i<g$.\smallskip
    
  For $i=g$ we have
   $$\Phi (S_{g, s}^g (\tau))=  S_{g-1, 2s}^{g-1}$$
   and for $i=0$ we have
    $$\Phi (S_{0, s}^g (\tau))= 2 S_{0, s}^{g-1}$$
   
    \end{lm} 
\noindent For $2^ks$ divisible by  $max\left( 16, \,2^g\right)$ we define
   $$ \Xi_{2^k s}^{(g)}=\sum_m \Xi_{2^k s}^{(g)}[m] =\frac{1}{2^g}\sum\limits_{i=0}^g (-1)^i 2^{\frac{i(i+1)}{2}}S_{i,s2^{k-i}}^g$$
  
   \noindent As immediate consequence we get
   \begin{thm}  For any $g$, let us   assume  that  $max\left( 16, \,2^g\right) $ divides $2^k s$, then
   $ \Xi_{2^k s}^{(g)}$ is a cusp form.
   \end{thm} 
{\it Proof}\, This is also an immediate consequence of the previous results, in fact we have
$$\Phi(  \Xi_{2^k s}^{(g)})=\frac{1}{2^g}\sum\limits_{i=0}^g (-1)^i 2^{\frac{i(i+1)}{2}}\Phi(S_{i,2^{k-i}s}^g)=$$
$$ \frac{1}{2^g}\sum\limits_{i=1}^{g-1} (-1)^i 2^{\frac{i(i+1)}{2}} (2^{i+1} S_{i, 2^{k-i}s}^{g-1}+S_{i-1, 2^{k-i+1}s}^{g-1})+$$

$$  \frac{(-1)^g }{2^g}  2^{\frac{g(g+1)}{2}} S_{g-1, 2^{k-g+1}s}^{g-1}=$$  

$$   \frac{1}{2^g}\left(\sum\limits_{i=2}^{g } (-1)^{i+1} 2^{\frac{i(i+1)}{2}}  S_{i-1, 2^{k-i+1}s}^{g-1}
+ \sum\limits_{i=1}^{g-1} (-1)^i 2^{\frac{i(i+1)}{2}} S_{i-1, 2^{k-i+1}s}^{g-1})  \right)+$$
$$ \frac{1}{2^g}\left( 2 S_{0, 2^{k}s}^{g-1}+(-1)^g  2^{\frac{g(g+1)}{2}} S_{g-1, 2^{k-g+1}s}^{g-1}\right)=0$$
Since we  get  different signs for the same  modular form.
\begin{rem} From a discussion with Sam Grushevsky, we realized that an alternative 
proof of this result, more in the spirit of this paper, is possible. According to Grushevsky's result we have that the modular forms
$\Xi_{2^k s}^{(g)}$ restricts on $\H_{g-1}\times\H_1$ as $\Xi_{2^k s}^{(g-1)}\cdot\Xi_{2^k s}^{(1)}$. \smallskip

\noindent So, applying the $\Phi$ operator is equivalent to compute
$$\lim_{ \lambda \rightarrow + \infty}\Xi_{2^k s}^{(g-1)}(\tau_1)\cdot\Xi_{2^k s}^{(1)}(i\lambda)=$$

$$\lim_{ \lambda \rightarrow + \infty} (S_{0, 2^ks}^{1}-2 S_{1, 2^{k-1}s}^{1})\Xi_{2^k s}^{(g-1)}(\tau_1)$$
 Now a simple  computation tells us that
 $$\lim_{ \lambda \rightarrow + \infty} (S_{0, 2^ks}^{1}-2 S_{1, 2^{k-1}s}^{1})=0$$
 \end{rem} 
As an immediate consequence of the previous theorem, we can reformulate the  result of Theorem 10 , in  fact we have  following
\begin{cor} For $g\leq 4$, the cosmological constant $\Xi^{(g)}=\Xi_{16}^{(g)}$ vanishes along $\M_g$
\end{cor}   
{\it Proof}  For $g\leq 4$,  $\Xi^{(g)}$ is a cusp form of weight 8. It is a well known fact that  $[\Gamma_g, 8]_0=0$ for $g\leq 3$.
Moreover the space $[\Gamma_4, 8]_0$ is one dimensional and it is generated by the Schottky polynomial $J(\tau)$, cf.\cite{HM}. \smallskip

\section{ Non existence of D'Hoker Phong form in genus 3}

As we  state in the introduction ,  Cacciatori, Dalla Piazza, and van Geemen, cf. \cite{CDPvG}  proposed an ansatz for the chiral superstring measure in genus 3 relaxing a  condition  posed by D' Hoker and Phong. We shortly explain why this relaxing is  necessary. This is not a proof.  A proof has been announced in   \cite{CDPvG} and it will appear in \cite{CG}.
  The solution proposed by D' Hoker and Phong should include a factor $\t_0^4$. This would imply that
  the proposed  ansatz for the superstring measure  should be
  $\t_0^4\Xi_6[0]$ with $\Xi_6[0]$ a modular form of weight 6 w.r.t $\Gamma_g(1,2)$ and a suitable character. This  will be the same character $\chi_g$  appearing in the transformation formula  of $\t_0^4$ , hence it is trivial when we restrict to $\Gamma_g(2)$
 Now when $g\leq 3$ the space of modular forms $[\Gamma_3(2), 6]$ is  generated by monomials in the theta constant,  that are particular theta series with characteristics.  The same is true for their subspaces.  In particular we have that theta series related to unimodular lattices of degree 12 are modular forms 
 in $[\Gamma_g(1,2), 6, \chi_g]$. Now according to the  table in \cite{CS} we have  three integral classes of such lattices, namely $I_{12}$, $E_8+I_4$ and $D_{12} ^+$, to these lattices ,up to multiplicative constants,  are associated the forms $F_1$, $F_3$ and $F_2$ in  the notation of \cite{CDPvG}.
 These forms are linearly independent already in genus $2$, thus we have
 $$ {\rm dim} [\Gamma_2(1,2), 6,  \chi_2]= {\rm dim}[\Gamma_3(1,2), 6,  \chi_3] =3$$
 Thus  simple computations will produce that in genus 3 we cannot have the desidered restriction properties.
 \section{Acknowledgements }
  I am very thankful to   Duong Phong, Sam Grushevsky and Bert van Geemen  for several helpful comments to a preliminary version of this manuscript.

\end{document}